\documentclass[pdflatex,sn-basic]{sn-jnl}

\usepackage[T1]{fontenc}
\usepackage{manyfoot}
\usepackage{graphicx}
\usepackage{amsmath,amssymb}
\usepackage{xcolor}
\usepackage{booktabs}
\usepackage{array}
\usepackage{tikz}
\usetikzlibrary{arrows.meta,positioning,shapes.geometric,calc,fit}

\definecolor{needred}{rgb}{0.80,0.00,0.00}

\begin{document}

\title[Hybrid Quantum-Classical Sequence Models for Calorimeter Energy Regression]{Hybrid Quantum-Classical Sequence Models for Sample-Efficient Energy Regression in Simulated Calorimeter Showers}

\author*[1]{\fnm{Tariq} \sur{Mahmood}}\email{tariqmahmood.chep@pu.edu.pk}
\author[1]{\fnm{Muhammad} \sur{Ahsan}}
\author[1]{\fnm{Talab} \sur{Hussain}}
\author[2]{\fnm{Grecia Itzel} \sur{Col\'in-Gonz\'alez}}
\author[3,4]{\fnm{Alfredo} \sur{Raya}}
\author[1]{\fnm{Muhammad Awais} \sur{Rafique}}
\author[5]{\fnm{HassabAlla M.A.} \sur{Mahmoud}}

\affil[1]{\orgdiv{Centre for High Energy Physics}, \orgname{University of the Punjab}, \city{Lahore}, \country{Pakistan}}
\affil[2]{\orgdiv{Facultad de Ciencias F\'isico-Matem\'aticas ``Mat. Luis Manuel Rivera Guti\'errez''}, \orgname{Universidad Michoacana de San Nicol\'as de Hidalgo}, \city{Morelia}, \country{Mexico}}
\affil[3]{\orgdiv{Facultad de Ingenier\'ia El\'ectrica}, \orgname{Universidad Michoacana de San Nicol\'as de Hidalgo}, \city{Morelia}, \country{Mexico}}
\affil[4]{\orgdiv{Centro de Ciencias Exactas}, \orgname{Universidad del B\'io-B\'io}, \city{Chill\'an}, \country{Chile}}
\affil[5]{\orgdiv{Department of Physics, College of Science}, \orgname{King Khalid University}, \orgaddress{\postcode{61413}, \city{Abha}, \country{Saudi Arabia}}}

\abstract{Energy reconstruction from electromagnetic calorimeter data underpins particle identification, event reconstruction and precision measurements in high-energy physics. Machine learning already performs well on this task, yet two questions remain open: how much the choice of data representation matters, and whether hybrid quantum-classical models add value in practice. We compare classical and quantum-enhanced models for incident positron energy reconstruction on the CaloGAN dataset, a public benchmark of 100{,}000 simulated electromagnetic shower events spanning 1 to 100~GeV. Eight architectures, namely SVR, ANN, CNN and LSTM together with their quantum counterparts, are trained on three calorimeter representations: engineered physical features, calorimeter images, and longitudinal shower sequences. Models that follow the longitudinal development of the shower reach the highest accuracy, which points to the value of physically informed representations. The hybrid QLSTM gives the lowest reconstruction error, with an RMSE of 0.294~GeV. Under matched-data conditions with only 1{,}500 training events, the QLSTM improves on the classical LSTM by 52.7\% and also surpasses an LSTM trained on the full set. These observations indicate that longitudinal representations are central to calorimeter energy reconstruction and give preliminary evidence that hybrid quantum-classical sequence models can be more sample-efficient for regression in high-energy physics.}


\maketitle

%======================================================================
\section{Introduction}\label{sec:intro}

High-energy physics seeks to describe the fundamental constituents of matter and the interactions that govern their behaviour. Modern collider experiments depend on detector systems that turn particle interactions into measurable signals, from which physical quantities are then reconstructed. Among those quantities, particle energy is one of the most consequential, since it feeds directly into particle identification, event reconstruction, precision measurements, and searches for physics beyond the Standard Model~\cite{griffiths2008,perkins2000}.

Electromagnetic calorimeters measure the energies of electrons, positrons and photons through the electromagnetic showers these particles produce inside dense absorbers~\cite{belayneh2020}. As a particle travels through the calorimeter, repeated bremsstrahlung and pair-production events build a cascade of secondary particles that deposit energy across several detector layers. The spatial and longitudinal distribution of those deposits carries information about the energy of the incident particle. Recovering that energy accurately is far from trivial, because shower fluctuations, detector effects and nonlinear correlations between deposited and incident energy all blur the relationship~\cite{behnke2013}.

Calorimeter reconstruction has traditionally relied on calibration procedures, analytical corrections and handcrafted observables. These methods work well, but the rising complexity of modern experiments and the growing volume of detector data have pushed the field towards machine learning~\cite{radovic2018}. Learning algorithms model complex nonlinear relationships directly from data and have performed strongly across particle physics, including event classification, jet tagging, particle identification, detector simulation and calorimeter reconstruction~\cite{baldi2014,belayneh2020}.

In parallel, progress in quantum computing has stimulated interest in quantum machine learning, a field that joins quantum information processing with learning algorithms~\cite{biamonte2017}. Hybrid quantum-classical architectures, which embed parameterized quantum circuits inside conventional learning pipelines, offer a way to test whether quantum representations help learning in concrete scientific problems. This work studies that question for calorimeter energy reconstruction.

\subsection{Related work}\label{sec:related}

Machine learning has become a standard tool in high-energy physics, and several reviews document its use across detector reconstruction, event classification, particle identification and data analysis~\cite{radovic2018,mondal2024}. Deep learning in particular has improved the extraction of information from complex detector signals.

For calorimeter applications, convolutional networks and other deep architectures have been used for particle reconstruction and energy estimation, often reaching accuracies beyond those of traditional methods~\cite{baldi2014,belayneh2020}. More recently, graph neural networks have been applied to particle tracking, particle-flow reconstruction and calorimeter event reconstruction, where they handle structured detector data well~\cite{ju2020,pata2021,qasim2022}.

Quantum machine learning has grown into an active area driven by the prospect of learning complex patterns through quantum representations~\cite{biamonte2017,havlicek2019,schuld2019}. Hybrid models have been proposed for classification and regression, including quantum kernel methods~\cite{havlicek2019,schuld2019}, variational quantum circuits~\cite{mitarai2018}, and quantum convolutional networks~\cite{cong2019}. Within high-energy physics the applications are recent but expanding: quantum approaches have been explored for jet classification, particle identification, track reconstruction and Higgs analyses~\cite{gianelle2022,fadol2024,duckett2024,bal2025}. Hybrid quantum-classical graph networks and quantum-enhanced classifiers report encouraging results on several benchmarks~\cite{tuysuz2021,felser2021}. Most of these studies, however, fix a single architecture or a single detector representation, which makes direct comparison between classical and quantum-enhanced approaches difficult.

Detector information can also be encoded in several ways, including engineered physical features, calorimeter images and sequential descriptions of shower development. Different architectures suit different encodings, yet few studies isolate the effect of the representation itself on reconstruction quality.

\subsection{Problem statement and contributions}\label{sec:gap}

Two gaps motivate this work. First, because most studies vary either the model or the representation but not both in a controlled way, it is hard to tell whether an observed gain comes from the architecture, from the way detector information is encoded, or from the interplay of the two. Second, the behaviour of hybrid quantum-classical models under limited-data conditions is poorly understood, even though scarce labelled data is common in scientific settings where simulation is expensive.

We address these gaps with a controlled comparison of classical and hybrid quantum-classical models for electromagnetic calorimeter energy reconstruction, using simulated positron showers from CaloGAN~\cite{paganini2018calogan,nachman2017data}. Eight learning architectures are evaluated on three complementary representations of the same events, which lets the effect of the representation be separated from the effect of the architecture. The study also probes performance when training data is deliberately reduced, so that sample efficiency can be measured under equal-data conditions. A further aim is to assess whether a hybrid sequential model offers a measurable benefit over its classical analogue, and to weigh the relative importance of representation, architecture and data availability. Rather than chasing a universal quantum advantage, the work provides a rigorous framework for understanding which factors most strongly shape calorimeter energy reconstruction.

%======================================================================
\section{Theoretical foundations}\label{sec:theory}

\subsection{High-energy physics detectors}\label{sec:detectors}

Collider experiments accelerate particles to high energies and bring them into controlled collisions. The resulting interactions produce many secondary particles whose properties must be measured precisely to reconstruct the underlying physics. Facilities such as the Large Hadron Collider at CERN rely on large detector systems built from concentric subsystems, each arranged around the beam axis and optimized for a particular measurement.

\begin{figure}
\centering
\includegraphics[width=0.7\textwidth]{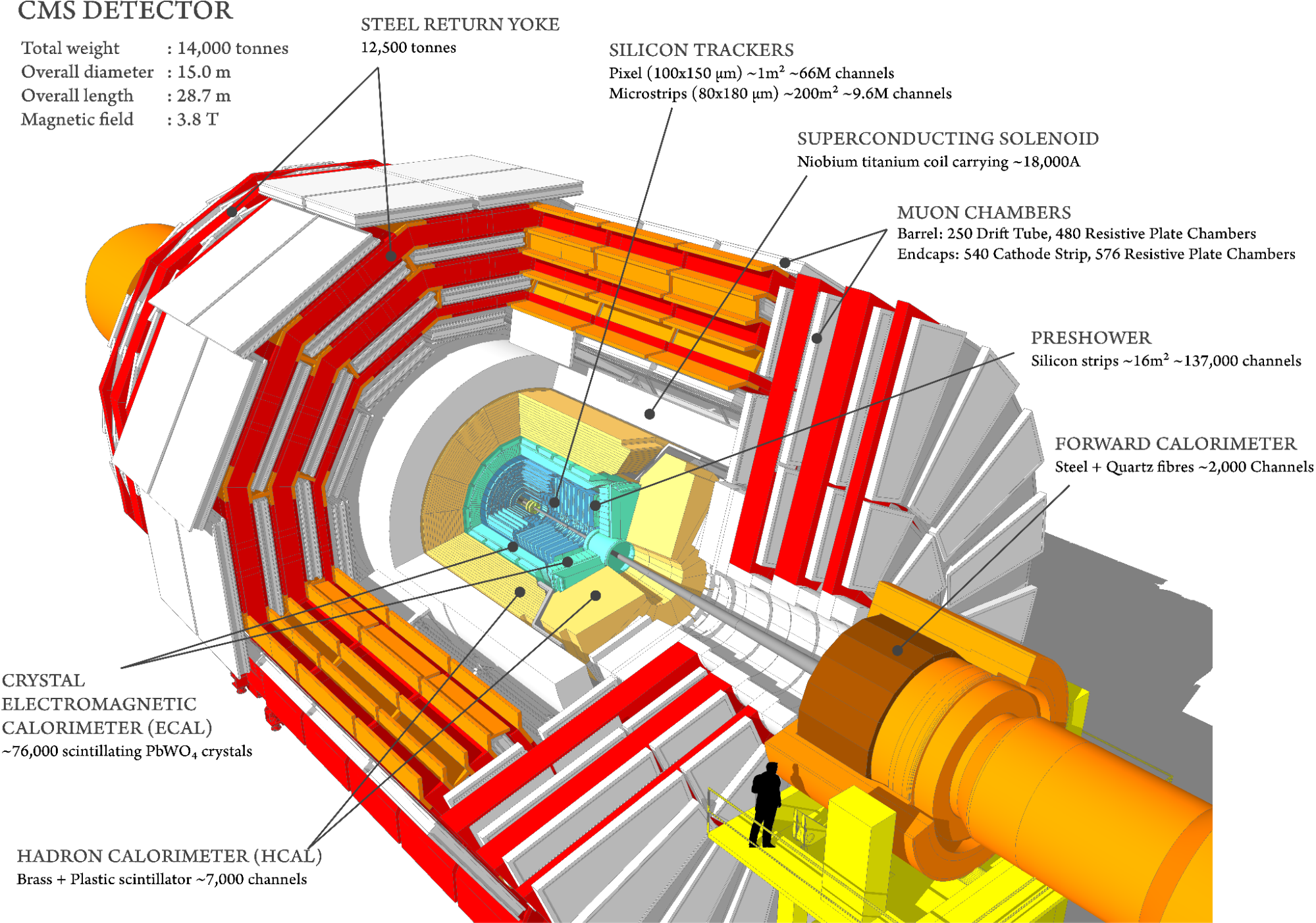}
\caption{
Sectional view of the CMS detector. The LHC beams travel in opposite directions along the central axis of the CMS cylinder and collide at the interaction point located at the detector center. The detector consists of multiple concentric subsystems including the silicon tracker, electromagnetic calorimeter (ECAL), hadronic calorimeter (HCAL), superconducting solenoid, and muon detection system. From the CMS Experiment at CERN. (https://cms.cern/news/cms-detector-design).}
\label{fig:moderndetector}
\end{figure}

At the centre lies the interaction point, where proton-proton collisions occur. Particles travel outward through successive layers, each adding complementary information. The innermost subsystem is the tracker, usually built from silicon sensors that record the trajectories of charged particles with fine spatial resolution; the curvature of those tracks inside a magnetic field yields momentum and charge. Surrounding the tracker is the electromagnetic calorimeter (ECAL), which measures the energies of electrons, positrons and photons through the showers they create. Because this study reconstructs positron energy from calorimeter data, the ECAL is described in more detail in Section~\ref{sec:ecal}. Outside the ECAL sits the hadronic calorimeter (HCAL), which measures strongly interacting particles such as pions, kaons and protons; hadronic showers are broader and more complex than electromagnetic ones. Many detectors add a superconducting solenoid whose field bends charged tracks for momentum measurement; in CMS the field reaches about 3.8~T. The outermost layer holds dedicated muon chambers, since muons traverse both calorimeters with little energy loss and are common in important final states. A sketch of the CMS detector is shown in Fig.~\ref{fig:moderndetector}.

The layered design produces complementary measurements that, combined through reconstruction algorithms, recover the properties of the particles created in the collision. Calorimeters are central to this picture because they provide the primary energy measurement; for electrons and positrons, the energy deposited in the ECAL is the foundation of the reconstruction procedures used throughout this work.

\subsection{Electromagnetic calorimetry}\label{sec:ecal}

Electromagnetic calorimeters estimate particle energy by absorbing it and measuring the response. When a high-energy electron or positron enters the calorimeter it radiates energetic photons through bremsstrahlung; those photons convert into electron-positron pairs through pair production; the new particles radiate again. This alternation builds an electromagnetic shower that deposits energy throughout the calorimeter volume, where the readout records it. The total deposited energy is proportional to the energy of the incident particle, so a suitable calibration recovers the original energy. This proportionality is the basis of electromagnetic calorimetry and of the reconstruction problem studied here.

Calorimeters interleave dense absorbers with active layers. The absorber drives shower development, while the active medium records the energy of the charged particles in the cascade. Common absorbers are lead and tungsten; active media include scintillating fibres, silicon sensors, liquid-argon chambers and crystal scintillators.

A central quantity in shower development is the radiation length $X_0$, the distance over which a high-energy electron loses, on average, a fraction $1/e$ of its energy to bremsstrahlung; for photons it is closely tied to the pair-production mean free path. Because these two processes dominate, $X_0$ sets the longitudinal scale of the shower. Calorimeter depth is usually quoted in radiation lengths, and modern devices span 20 to 30~$X_0$ to contain the shower almost fully, since energy that leaks beyond the detector biases the reconstruction.

Two families exist. In homogeneous calorimeters a single material acts as both absorber and active medium, so the whole shower contributes to the signal; the lead-tungstate crystals of the CMS ECAL at CERN are a representative example. In sampling calorimeters, separate absorber and active layers alternate; only a fraction of the shower is sampled, but the design gains in cost, scalability and radiation tolerance.

The quality of the energy measurement is captured by the relative energy resolution, conventionally written as a sum in quadrature of three terms,
\begin{equation}\label{eq:resolution}
\frac{\sigma(E)}{E} \;=\; \frac{a}{\sqrt{E}} \;\oplus\; \frac{b}{E} \;\oplus\; c
\;=\; \sqrt{\left(\frac{a}{\sqrt{E}}\right)^{2} + \left(\frac{b}{E}\right)^{2} + c^{2}},
\end{equation}
where $\sigma(E)$ is the uncertainty on the reconstructed energy, $E$ is the incident energy, $a$ is the stochastic term from intrinsic shower fluctuations, $b$ collects electronic and detector noise, and $c$ is the constant term from calibration and systematic effects. %\needinfo{Before we had $a/E^2 + b/E^2 + c^2$, which is dimensinally wrong. Can you confirm this standard expression, please?}

\begin{figure}
\centering\includegraphics[width=0.7\textwidth]{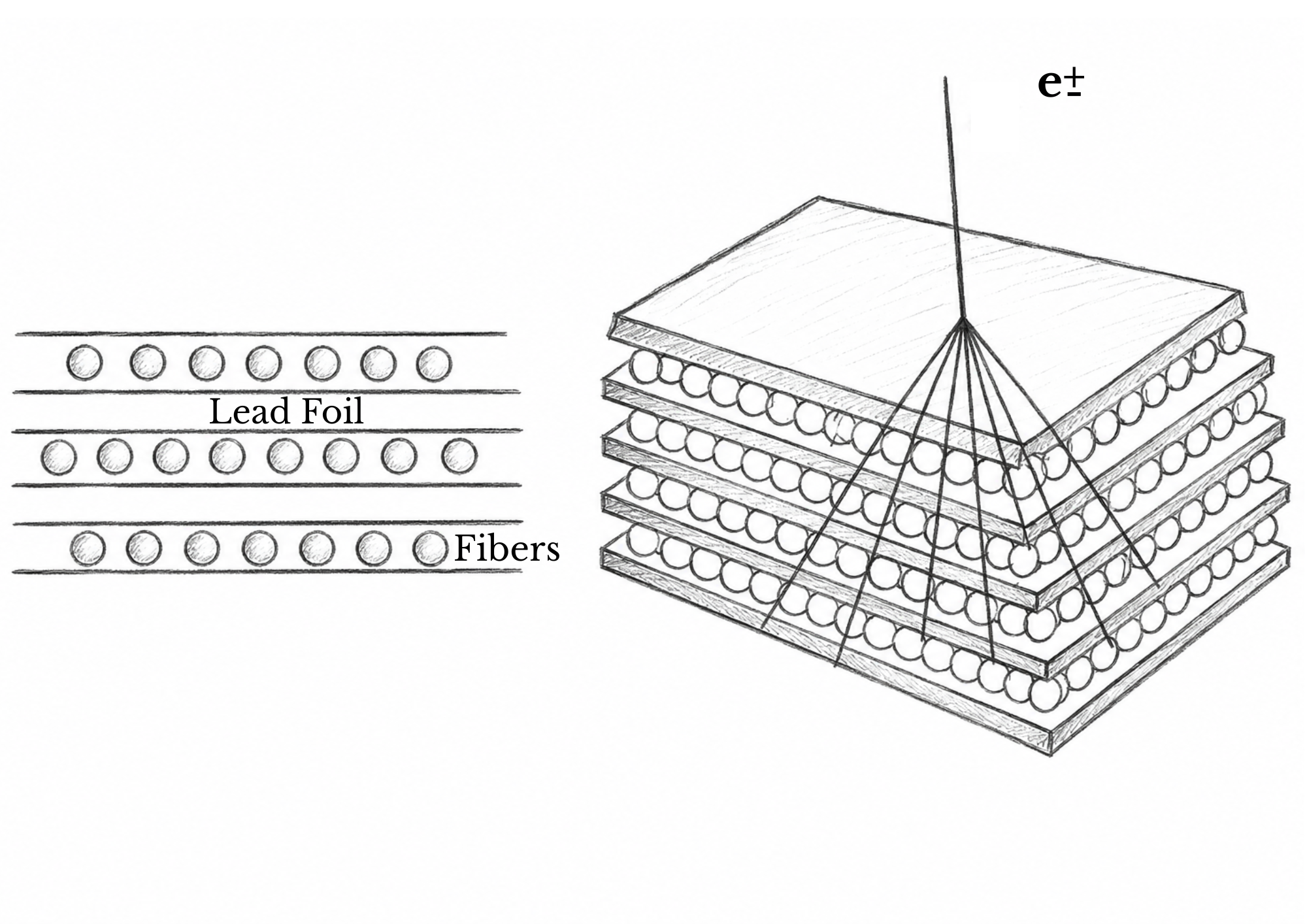}
\caption{Schematic of an electromagnetic calorimeter: shower development through alternating lead absorber and scintillating-fibre layers,  resolution curve and an ECAL module}\label{fig:ecal}
\end{figure}

For electromagnetic calorimeters, as sketched in Fig.~\ref{fig:ecal}, the stochastic term usually dominates over a broad range, so the resolution approaches
\begin{equation}\label{eq:stochastic}
\frac{\sigma(E)}{E} \;\propto\; \frac{1}{\sqrt{E}},
\end{equation}
which means the relative resolution improves as the energy grows. This scaling is a defining feature of electromagnetic calorimeters and serves later as a reference for interpreting the machine learning results.

Although total deposited energy is the primary observable, a calorimeter records much more. Energy spreads across many cells and layers, producing spatial and longitudinal patterns that reflect the shower development. These patterns can be expressed as engineered features, calorimeter images, or longitudinal profiles, and because each form preserves a different aspect of the cascade, they provide a natural basis for the machine learning approaches used here.

\subsection{Electromagnetic showers}\label{sec:showers}

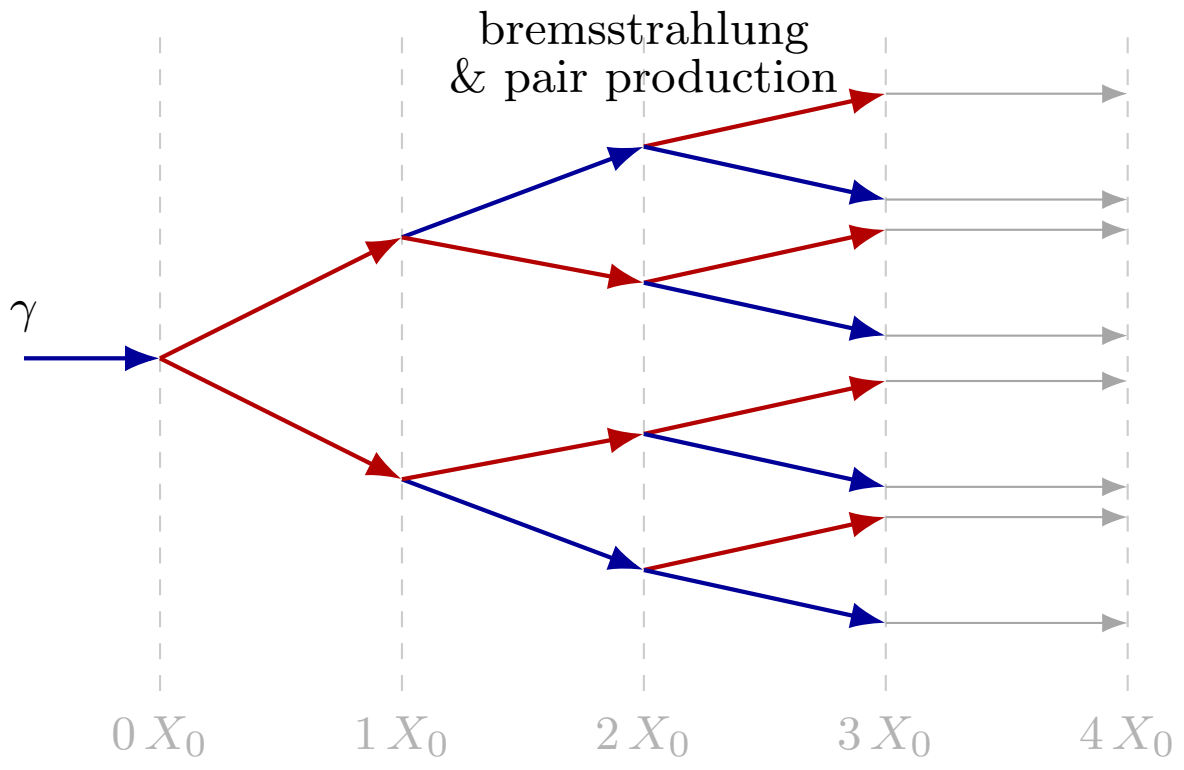
\begin{figure}[t]
\centering
\begin{tikzpicture}[scale=1.0,>=Latex,
  ph/.style={->,thick,blue!60!black},
  el/.style={->,thick,red!70!black}]
  \foreach \x/\lab in {0/0,1.6/1,3.2/2,4.8/3,6.4/4}{
    \draw[gray!40,dashed] (\x,-2.2) -- (\x,2.2);
    \node[gray!60,font=\scriptsize] at (\x,-2.5) {$\lab\,X_0$};
  }
  % generation 0: incident photon
  \draw[ph] (-0.9,0) -- (0,0);
  \node[font=\scriptsize] at (-0.9,0.28) {$\gamma$};
  % gen 0 -> 1 pair
  \draw[el] (0,0) -- (1.6,0.8);
  \draw[el] (0,0) -- (1.6,-0.8);
  % gen 1 -> 2 brems
  \draw[ph] (1.6,0.8) -- (3.2,1.4);
  \draw[el] (1.6,0.8) -- (3.2,0.5);
  \draw[ph] (1.6,-0.8) -- (3.2,-1.4);
  \draw[el] (1.6,-0.8) -- (3.2,-0.5);
  % gen 2 -> 3
  \foreach \y in {1.4,0.5,-0.5,-1.4}{
    \draw[el] (3.2,\y) -- (4.8,{\y+0.35});
    \draw[ph] (3.2,\y) -- (4.8,{\y-0.35});
  }
  % gen 3 -> 4 (fan)
  \foreach \y in {1.75,1.05,0.85,0.15,-0.15,-0.85,-1.05,-1.75}{
    \draw[gray!70,->] (4.8,\y) -- (6.4,{\y});
  }
  \node[font=\scriptsize,align=center] at (3.2,2.0) {bremsstrahlung\\ \& pair production};
\end{tikzpicture}
\caption{Schematic development of an electromagnetic shower in an absorber. An incident photon initiates the cascade through pair production; the secondary electrons and positrons radiate bremsstrahlung photons that convert into further pairs, so the particle count grows with depth until ionization losses halt the multiplication. Depth is shown in units of the radiation length $X_0$.}
\label{fig:shower}
\end{figure}

Electromagnetic showers are cascades produced when high-energy electrons, positrons or photons interact electromagnetically with matter. For an incident electron or positron the dominant process is bremsstrahlung, written schematically as
\begin{equation}
e^{\pm} \rightarrow e^{\pm} + \gamma,
\end{equation}
while for an incident photon the shower starts through pair production,
\begin{equation}
\gamma \rightarrow e^{+} + e^{-}.
\end{equation}
Each emitted photon can start a new pair, and each new electron or positron radiates further, so the particle count multiplies as the cascade advances, as sketched in Fig.~\ref{fig:shower}. The number of particles grows quickly at first, then the average energy per particle drops until ionization losses take over, the multiplication stops, and the remaining particles deposit their energy.

A simple picture follows from the Heitler model, in which every particle splits into two of roughly equal energy after each radiation length. After $t$ radiation lengths the particle count and the mean energy per particle are
\begin{equation}\label{eq:heitler}
N(t) = 2^{\,t}, \qquad E(t) = \frac{E_0}{2^{\,t}},
\end{equation}
where $E_0$ is the energy of the primary particle. Multiplication continues until the mean energy falls to the critical energy $E_c$, below which ionization losses exceed radiative losses. The shower maximum occurs when $E(t)\approx E_c$, which gives the depth of maximum development,
\begin{equation}\label{eq:tmax}
t_{\max} \;\approx\; \frac{\ln(E_0/E_c)}{\ln 2}.
\end{equation}
%\needinfo{The draft omited the factor $1/\ln 2$ in $t_{\max}$ and, in the paragraph, critical energy was labeled as $E_0$; in this version, $E_c$  is used consistently. Please, verify}

Equation~\eqref{eq:tmax} shows that higher-energy particles penetrate deeper before reaching maximum deposition, so the longitudinal profile of the shower encodes the incident energy. The deposited energy is not uniform: characteristic longitudinal and transverse profiles emerge, the first describing how deposition evolves with depth and the second the lateral spread around the shower axis. Both follow from the stochastic nature of bremsstrahlung and pair production, and both carry more information than the total energy alone. The three representations used in this work, namely engineered features, calorimeter images and longitudinal sequences, are built precisely from these patterns.

\subsection{Energy reconstruction in calorimeters}\label{sec:reco}

The goal of a calorimeter is to estimate the energy of the incident particle, not merely to record deposits. In an ideal device the total deposited energy is exactly proportional to the incident energy, so the reconstructed value is
\begin{equation}\label{eq:linreco}
E_{\mathrm{reco}} = C \cdot S,
\end{equation}
with $S$ the measured signal and $C$ a calibration constant. In practice, shower fluctuations, incomplete containment, noise, calibration uncertainty, non-uniform response and leakage all break this simple linearity, which is why accurate reconstruction needs more than a single proportionality.

A calorimeter event is a collection of energy deposits $E_{ij}^{(l)}$, where $i,j$ index the cell within layer $l$. From this high-dimensional response several physically meaningful quantities follow. The total deposited energy is
\begin{equation}\label{eq:etot}
E_{\mathrm{tot}} = \sum_{l,i,j} E_{ij}^{(l)},
\end{equation}
the energy in layer $l$ and its fraction are
\begin{equation}\label{eq:elayer}
E_l = \sum_{i,j} E_{ij}^{(l)}, \qquad f_l = \frac{E_l}{E_{\mathrm{tot}}},
\end{equation}
and the transverse centroid and width along a coordinate $x$ are
\begin{equation}\label{eq:moments}
\bar{x} = \frac{\displaystyle{\sum_i E_i\, x_i}}{\displaystyle{\sum_i E_i}}, \qquad
\sigma_x = \sqrt{\frac{\displaystyle{\sum_i E_i\,(x_i-\bar{x})^2}}{\displaystyle{\sum_i E_i}}}.
\end{equation}
The layer fractions describe how energy is distributed along the calorimeter depth, while the moments characterize the lateral spread; both correlate with incident energy.

Historically these handcrafted observables, together with calibration constants, drove the reconstruction. As detectors produce higher-dimensional data, machine learning offers an alternative that learns directly from the response and can exploit spatial, longitudinal and statistical structure. Depending on the encoding, the same event becomes a vector of engineered features, an image-like map, or an ordered longitudinal sequence, which sets up the comparison pursued in this study.

\subsection{Machine learning for energy reconstruction}\label{sec:ml}

Machine learning frames reconstruction as a function learned from data rather than fixed in advance. For calorimetry the objective is a mapping from the detector response to the true incident energy. Formally this is supervised regression: given an input vector $\mathbf{x}$ built from the calorimeter response, the model approximates
\begin{equation}\label{eq:regression}
\hat{E} = f(\mathbf{x}),
\end{equation}
where $\hat{E}$ is the reconstructed energy and $\mathbf{x}$ may hold engineered features, calorimeter images, longitudinal profiles, or combinations of these.

The strength of machine learning here comes from its ability to model nonlinear relationships between detector measurements and particle energy. Shower development is shaped by stochastic interactions, geometry, leakage and deposition fluctuations, which together create patterns that resist simple analytical description. Several paradigms suit the problem. Support vector regression operates on compact engineered observables and performs well when informative features are available. Artificial neural networks learn nonlinear combinations of detector variables through trainable layers. Convolutional networks read spatial correlations from calorimeter images. Recurrent architectures such as long short-term memory networks process sequences and naturally model the longitudinal evolution of the shower across layers.

A point that runs through this work is that the encoding matters as much as the algorithm. Engineered features emphasize physically motivated observables, images preserve spatial structure, and longitudinal sequences capture how deposition evolves with depth. The effectiveness of a model therefore depends on both the architecture and the representation it receives.

\subsection{Quantum machine learning}\label{sec:qml}

Quantum machine learning joins quantum computing with learning algorithms, with the aim of exploiting superposition, interference and entanglement to represent data and compute in ways that are hard to reproduce classically. Where classical computers store information in bits, quantum computers use qubits, and a single qubit can occupy a superposition
\begin{equation}\label{eq:qubit}
|\psi\rangle = \alpha|0\rangle + \beta|1\rangle, \qquad |\alpha|^2 + |\beta|^2 = 1,
\end{equation}
with complex amplitudes $\alpha$ and $\beta$. Combining $n$ qubits gives access to a Hilbert space of dimension $2^{n}$, which motivates quantum representations for high-dimensional problems. Entanglement adds correlations that cannot be written as independent combinations of single-qubit states and is often regarded as a primary source of expressive power.

Most practical algorithms today are hybrid: classical computers handle optimization and parameter updates, while quantum circuits act as trainable layers that map data into quantum feature spaces. This split matters in the present noisy intermediate-scale quantum era, where hardware is limited in qubit count, circuit depth and noise tolerance. A typical hybrid model encodes classical data into a quantum state, transforms it with parameterized gates, and measures the result, after which classical optimization closes the loop. A parameterized quantum circuit can be written as
\begin{equation}\label{eq:pqc}
f(\mathbf{x};\theta) = \langle 0 |\, U^{\dagger}(\mathbf{x})\, W^{\dagger}(\theta)\, \hat{O}\, W(\theta)\, U(\mathbf{x})\, | 0 \rangle,
\end{equation}
where $U(\mathbf{x})$ encodes the input, $W(\theta)$ is the trainable variational circuit, and $\hat{O}$ is the measured observable whose expectation value is the model output.

A widely used idea is the quantum feature space, in which classical inputs are mapped to quantum states and similarities are computed from state overlaps. For two inputs $\mathbf{x}_i$ and $\mathbf{x}_j$ the quantum kernel is
\begin{equation}\label{eq:kernel}
K_Q(\mathbf{x}_i,\mathbf{x}_j) = \big| \langle \phi(\mathbf{x}_i) | \phi(\mathbf{x}_j) \rangle \big|^{2},
\end{equation}
with $|\phi(\mathbf{x})\rangle$ the encoded state; such kernels underpin quantum support vector methods~\cite{havlicek2019,schuld2019}. A complementary family uses variational quantum circuits as trainable models~\cite{mitarai2018}, which fit neatly into neural-network-like pipelines for supervised learning.

Whether quantum methods offer a practical advantage is still open. Noise, short coherence times and shallow circuits limit current hardware, so many studies, including this one, evaluate quantum models on statevector simulators under idealized conditions before moving to physical devices. In high-energy physics the interest is driven by the volume of detector data and the intricate correlations within it, with applications reported in event classification, jet tagging, anomaly detection, track reconstruction and calorimeter reconstruction~\cite{gianelle2022,duckett2024,tuysuz2021}. This work evaluates four representative approaches, namely a quantum kernel regressor, a quantum neural network, a quantum convolutional network and a quantum long short-term memory network, which together span kernel-based, variational, convolutional and sequential paradigms.

%======================================================================
\section{Methodology}\label{sec:method}

\subsection{Dataset and event selection}\label{sec:dataset}

The study uses the CaloGAN dataset~\cite{paganini2018calogan,nachman2017data}, a public benchmark that reproduces the response of a high-granularity electromagnetic calorimeter through detailed Monte Carlo simulation and has become a standard reference for machine learning studies in calorimetry. The sample contains positron-induced electromagnetic showers with incident energies between 1 and 100~GeV. As each positron crosses the calorimeter it produces a cascade that deposits energy across the detector layers, and the resulting pattern encodes both the incident energy and the development of the shower, which makes the data well suited to supervised energy regression.

A total of 100{,}000 events was used, each labelled with its known incident energy. The data were split once, with a fixed seed (\texttt{SEED}~$=42$) and shuffling, into training, validation and test subsets following Table~\ref{tab:partition}; the same index split was reused across all classical and quantum models so that every architecture sees identical events. For the classical models the regression target is the true incident energy in GeV (from 1.0003 to 99.9967~GeV). For the quantum models the target was standardized to zero mean and unit variance with a \texttt{StandardScaler} fitted on the training subset and inverse-transformed back to GeV for evaluation.

\begin{table}[t]
\centering
\caption{Dataset partition used for model development and evaluation.}
\label{tab:partition}
\begin{tabular}{lcc}
\toprule
Subset & Events & Percentage \\
\midrule
Training   & 70{,}000  & 70\% \\
Validation & 15{,}000  & 15\% \\
Test       & 15{,}000  & 15\% \\
\midrule
Total      & 100{,}000 & 100\% \\
\bottomrule
\end{tabular}
\end{table}

Before any split or scaling, the data were checked for negative and non-numeric values across the energy target and the three calorimeter layers as well as the overflow region; none were found, and all features and targets were stored in single precision. Preprocessing artefacts such as scalers, principal-component objects, split indices and metadata were saved to disk for full reproducibility.

\begin{figure}
\centering
\includegraphics[width=0.9\textwidth]{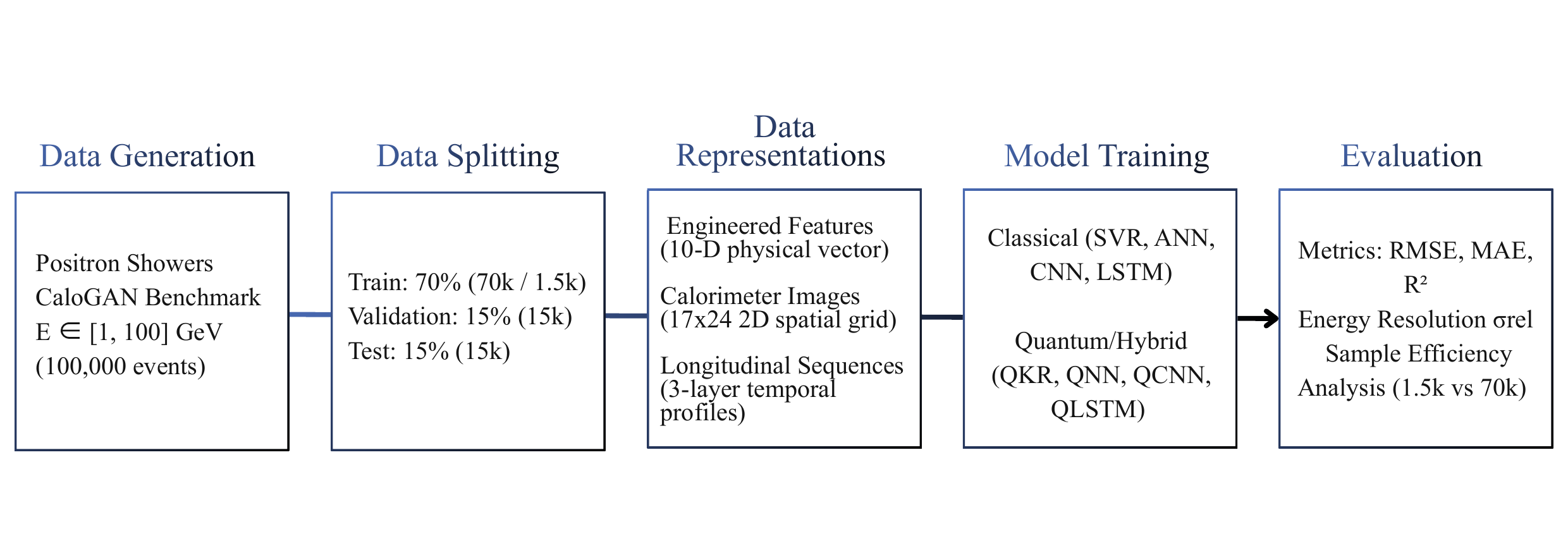}
\caption{Data-processing workflow: positron generation, shower simulation, detector readout, train/validation/test split, preprocessing, construction of engineered features, calorimeter images and longitudinal sequences, model training and evaluation.}\label{fig:dataflow}
\end{figure}

Because one aim of the work is to measure the effect of training-set size, a reduced subset of 1{,}500 training events was also drawn from the full training split, with matched validation and test events, to study sample efficiency under equal-data conditions. The reduced-data protocol applies identical subsets, preprocessing and metrics to every participating model, so that any difference can be attributed to the architecture rather than to data volume. The data-processing workflow is depicted in Fig.~\ref{fig:dataflow}.

\subsection{Calorimeter data representations}\label{sec:repr}

\begin{figure}
\centering
\includegraphics[width=0.50\linewidth]{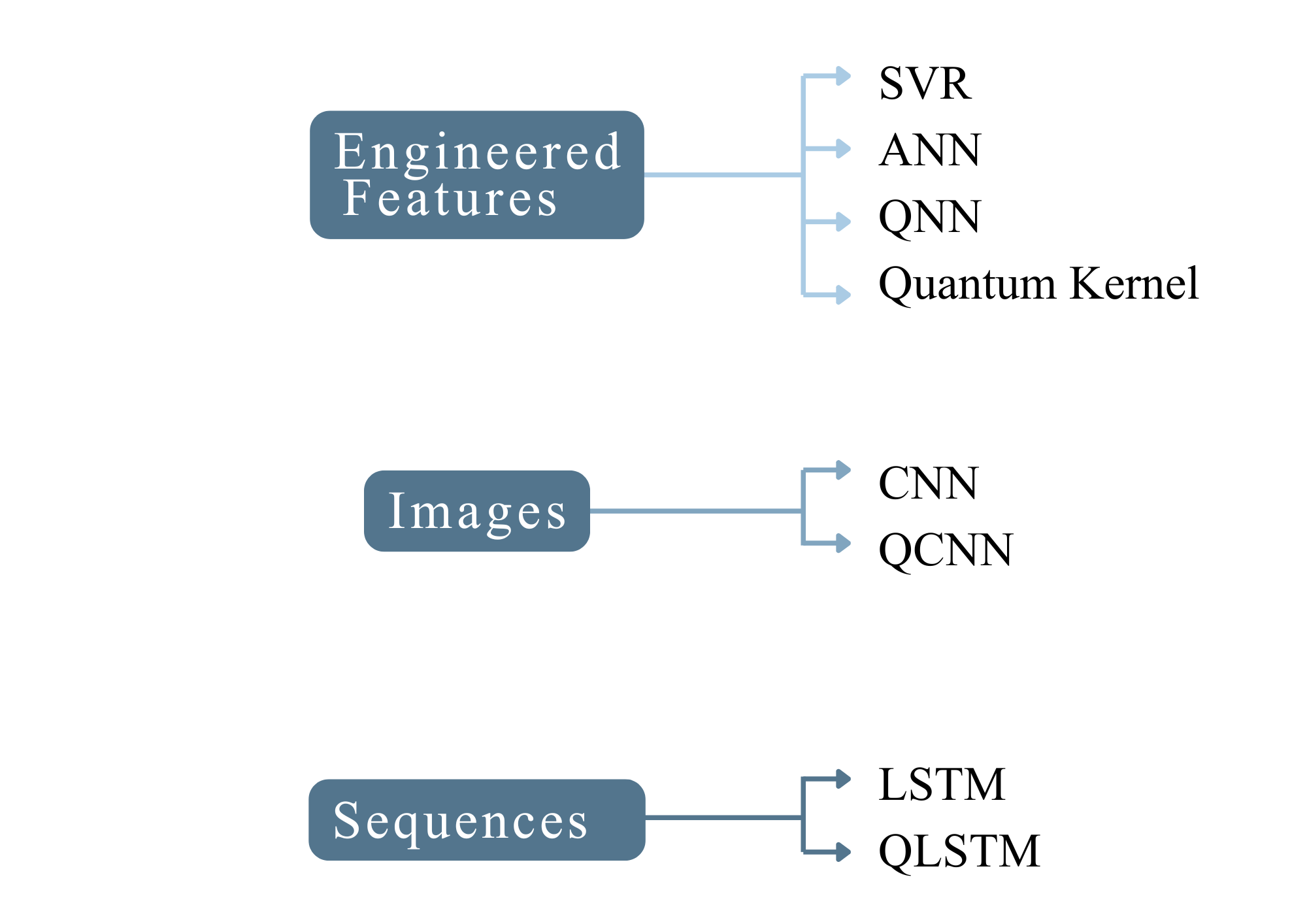}
\caption{Mapping between the calorimeter data representations and the learning architectures evaluated in this study. Engineered features are processed by SVR, ANN, QNN and the quantum kernel regressor; calorimeter images by CNN and QCNN; and longitudinal shower sequences by LSTM and QLSTM. This organization provides the experimental framework used to compare the influence of data representation and learning architecture.}
\label{fig:model_map}
\end{figure}

The same event was cast into three complementary representations, each preserving a different aspect of the shower so that the influence of the encoding can be separated from that of the architecture, as summarized in Fig.~\ref{fig:model_map}.

\paragraph{Engineered physical features.} Nine handcrafted observables summarize the energy scale, the per-layer signal and the longitudinal profile, as listed in Table~\ref{tab:features}. The nine-dimensional vectors were standardized with a \texttt{StandardScaler} fitted on the training set, and a small constant ($\varepsilon=10^{-8}$) stabilized the fractional features. This representation feeds the SVR, the engineered-feature ANN, the quantum kernel regressor and the QNN. The single \texttt{total\_deposited} feature also serves as the sole input to a linear-regression baseline.

\begin{table}[t]
\centering
\caption{Engineered calorimeter features.}
\label{tab:features}
\begin{tabular}{p{0.24\linewidth} p{0.66\linewidth}}
\toprule
Feature & Meaning \\
\midrule
\texttt{layer0\_sum}--\texttt{layer2\_sum} & Energy deposited in each calorimeter layer (GeV) \\
\texttt{overflow\_sum} & Energy outside the selected calorimeter region (GeV) \\
\texttt{total\_deposited} & Sum over the three layers and the overflow (GeV) \\
\texttt{frac\_l0}--\texttt{frac\_l2} & Fraction of total energy in each layer \\
\texttt{frac\_overflow} & Fraction of total energy in the overflow region \\
\bottomrule
\end{tabular}
\end{table}

\paragraph{Calorimeter images.} The raw energy deposits were kept on the native detector grids, of size $3\times96$, $12\times12$ and $12\times6$ for layers~0, 1 and 2, with no resizing, padding or geometric harmonization. For the classical CNN a singleton channel was added to give tensors of shape $(N,3,96,1)$, $(N,12,12,1)$ and $(N,12,6,1)$, with no intensity or target normalization, and the three layers were processed by three parallel convolutional branches. For the hybrid QCNN each event was summarized by eight statistics, namely the per-layer mean and standard deviation for the three layers together with the layer~0 and layer~1 sums, and these were rescaled to $[-\pi,\pi]$ with a \texttt{MinMaxScaler} fitted on the QCNN training subset before angle encoding.

\paragraph{Longitudinal shower sequences.} Each event was written as an ordered sequence of length three along the shower direction (layer~0~$\rightarrow$~layer~1~$\rightarrow$~layer~2). Each step carries two quantities, the total deposited energy and its fraction of the event total, giving a tensor of shape $(N,3,2)$. The sequence features were standardized with a \texttt{StandardScaler} fitted on the training set, reshaping to two dimensions and back. This representation feeds the LSTM and the QLSTM.

For the compressed inputs of the vector-based quantum models, the standardized 504-dimensional flattened image vector was reduced with principal-component analysis to eight components (PCA-8, retaining 22.9\% of the variance) and to sixteen components (PCA-16, 25.5\%), with the PCA fitted on the training data only and the components rescaled to $[-\pi,\pi]$ to match the angle encoding. The QCNN does not use PCA; only the QNN and the quantum kernel rely on it.

\subsection{Classical models}\label{sec:classical}

Four classical paradigms provide the baselines. Support vector regression~\cite{cortes1995,scholkopf2002} was trained on the nine engineered features with a radial-basis kernel, regularization $C=10$, $\varepsilon=0.1$ and \texttt{gamma}~$=$~\texttt{scale}. An artificial neural network on the same features used two hidden layers of 64 and 32 units with ReLU activations and dropout~0.1, for 2{,}753 trainable parameters; a second ANN on the 504-dimensional flattened image used hidden layers of 256 and 128 units with dropout~0.2, for 162{,}305 parameters. A multi-branch convolutional network read the three calorimeter images through independent streams of two convolutions each, concatenated before a dense regression head, for 555{,}889 parameters. A long short-term memory network~\cite{hochreiter1997} read the longitudinal sequences with a single recurrent layer of 64 units followed by a dense layer of 32 units, for 19{,}265 parameters. The engineered-feature ANN and both LSTM variants were trained with three seeds (42, 123, 999) to estimate the statistical spread.

\subsection{Quantum and hybrid models}\label{sec:quantum}

Four quantum approaches mirror the classical ones so that the architecture can be compared at a fixed representation. A quantum kernel regressor replaces the classical kernel with state overlaps computed from an angle-encoded feature map and feeds the resulting matrix to a classical optimizer; its $O(N^2)$ kernel cost limited training to 300 events. A quantum neural network encodes the PCA-reduced features into an eight-qubit variational circuit. A quantum convolutional network applies variational operations to the eight image summary statistics. A quantum long short-term memory network~\cite{chen2022qlstm} embeds variational circuits inside the recurrent cell and reads the longitudinal sequences, which aligns the model with the ordered structure of shower development.

All quantum and hybrid models were implemented with PennyLane~\cite{bergholm2018} together with PyTorch or scikit-learn under ideal statevector simulation. A common configuration was kept where possible: angle embedding with rotations about the $Y$ axis and strongly entangling layers as the variational ansatz, on eight qubits, as summarized in Table~\ref{tab:qspecs}. 
The total number of trainable parameters in the hybrid quantum models is modest. The QNN\_PCA8 model contains 2,761 trainable parameters, comprising 72 quantum variational parameters and 2,689 classical parameters. The QCNN\_Hybrid model contains 393 trainable parameters (72 quantum and 321 classical). The QLSTM\_Hybrid model contains 1,585 trainable parameters, including 72 quantum parameters and 1,513 classical parameters from the recurrent and fully connected layers.
%\needinfo{the number of trainable parameters for  QNN, QCNN and QLSTM are missing in the draft. Can you please provide these numbers?}

\begin{table}[t]
\centering
\caption{Quantum circuit specifications (verified through PennyLane circuit inspection). All models use eight qubits and statevector simulation. SEL denotes strongly entangling layers; AE($R_y$) denotes angle embedding with $Y$ rotations.}
\label{tab:qspecs}
\begin{tabular}{lccc}
\toprule
Model & Qubits & Encoding / ansatz & Inputs \\
\midrule
Quantum kernel & 8 & AE($R_y$) fidelity kernel & PCA features \\
QNN            & 8 & SEL + AE($R_y$)          & PCA-8 features \\
QCNN (hybrid)  & 8 & SEL + AE($R_y$)          & 8 image statistics \\
QLSTM (hybrid) & 8 & SEL + AE($R_y$)          & longitudinal sequence \\
\bottomrule
\end{tabular}
\end{table}

\subsection{Training procedure and evaluation metrics}\label{sec:training}

All deep models were trained by supervised regression on the true incident energy with the Adam optimizer, a learning rate of $10^{-3}$, mean-squared-error loss, a batch size of 256 for the full dataset and 64 for the matched subset, up to 100 epochs with early stopping on the validation loss and learning-rate reduction on plateau. The principal hyperparameters are collected in Table~\ref{tab:hyper}, reconstructed from the model descriptions above. A linear regression on the total deposited energy provides a trivial physics baseline that any useful model must beat, since the total deposited energy is already nearly proportional to the incident energy.

\begin{table}[t]
\centering
\caption{Training configuration, reconstructed from the model descriptions. All deep models use Adam with learning rate $10^{-3}$, mean-squared-error loss, up to 100 epochs with early stopping, and learning-rate reduction on plateau.}% \needinfo{the counting of parameters in the quantum models is yet to be confirmed}}
\label{tab:hyper}
\begin{tabular}{lccc}
\toprule
Model & Representation & Batch & Trainable params \\
\midrule
SVR              & engineered (9)   & --   & non-parametric \\
ANN (engineered) & engineered (9)   & 256  & 2{,}753 \\
ANN (flattened)  & image (504)      & 256  & 162{,}305 \\
CNN (multibranch)& images           & 256  & 555{,}889 \\
LSTM             & sequence (3,2)   & 256  & 19{,}265 \\
QNN              & PCA-8            & --   & 2{,}761 \\
QCNN (hybrid)    & 8 statistics     & --   & 393 \\
QLSTM (hybrid)   & sequence (3,2)   & 64   & 1{,}585 \\
\bottomrule
\end{tabular}
\end{table}

Performance was measured with five complementary metrics. The mean absolute error and the root-mean-square error are
\begin{equation}\label{eq:mae}
\mathrm{MAE} = \frac{1}{n}\sum_{i=1}^{n}\big|\hat{E}_i - E_i\big|,
\qquad
\mathrm{RMSE} = \sqrt{\frac{1}{n}\sum_{i=1}^{n}\big(\hat{E}_i - E_i\big)^2},
\end{equation}
where $E_i$ and $\hat{E}_i$ are the true and reconstructed energies. The coefficient of determination and the mean relative error are
\begin{equation}\label{eq:r2}
R^{2} = 1 - \frac{\displaystyle{\sum_i (\hat{E}_i - E_i)^2}}{\displaystyle{\sum_i (E_i - \bar{E})^2}},
\qquad
\mathrm{MRE} = \frac{1}{n}\sum_{i=1}^{n}\frac{\big|\hat{E}_i - E_i\big|}{E_i},
\end{equation}
with $\bar{E}$ the mean true energy. Finally, the relative energy resolution is the standard deviation of the relative error,
\begin{equation}\label{eq:sigmarel}
\sigma_{\mathrm{rel}} = \mathrm{std}\!\left(\frac{\hat{E}-E}{E}\right),
\end{equation}
which connects the learning metrics to the calorimetric quantity of Eq.~\eqref{eq:resolution}. RMSE and $\sigma_{\mathrm{rel}}$ are emphasized because calorimeter reconstruction is sensitive to rare large errors. Models trained with several seeds are reported as mean and standard deviation. 
We have repeated the QLSTM training using three random seeds (42, 123, 999), consistent with the classical LSTM protocol. The model was trained on the matched subset of 1,500 events (with 300 validation and 300 test samples). The resulting performance is $RMSE = 0.271 \pm 0.002$ and $MAE = 0.128 \pm 0.001$ (mean $\pm$ standard deviation across seeds). The seed-to-seed variation is small, demonstrating statistical stability of the hybrid QLSTM model. Table 6 has been updated accordingly.
%\needinfo{QLSTM is quoted only with a single seed. Can you please evaluate for other seeds as well? If not, can you explain explicitly the cost of the simulation?}

\subsection{Sample-efficiency evaluation}\label{sec:sampleeff}

To isolate the effect of architecture from that of data volume, a matched-subset protocol trained every participating model on the same 1{,}500 events, with matched validation and test sets drawn with \texttt{SEED}~$=42$. The comparison centers on the LSTM and the QLSTM, since both read the same longitudinal representation and differ mainly in the quantum circuits. Table~\ref{tab:matched} summarizes the matched-subset results.

\begin{table}[t]
\centering
\caption{Matched-subset comparison ($N=1{,}500$ for the matched column). Percentages give the RMSE increase relative to the full-data model.}
\label{tab:matched}
\begin{tabular}{lcccc}
\toprule
Model & $N$ (full) & RMSE full & RMSE matched & Change \\
\midrule
SVR              & 70k   & 0.799        & $\sim$1.02        & $+27.7\%$ \\
ANN (engineered) & 70k   & $0.442\pm0.040$ & $\sim0.68\pm0.05$ & $+53.8\%$ \\
LSTM             & 70k   & $0.429\pm0.052$ & $0.621\pm0.057$   & $+44.8\%$ \\
QLSTM (hybrid)   & 1.5k & $0.271 \pm 0.002$ & \textbf{0.294} & best at 1.5k \\
\bottomrule
\end{tabular}
\end{table}

%\needinfo{QLSTM row mixes contradictory figures: the rest of the manuscript trains the QLSTM with  1{,}500 events and  reports RMSE 0.294, but the table indicated  N(full)$=2{,}500$ with RMSE 0.446$\pm$0.052, which is almost identical to the full LSTM. Can you please confirm the size of the training data and if necessary, correct the row?}

%======================================================================
\section{Results and discussion}\label{sec:results}

\subsection{Dataset characterization}\label{sec:datachar}

\begin{figure}
\centering
\includegraphics[width=0.7\textwidth]{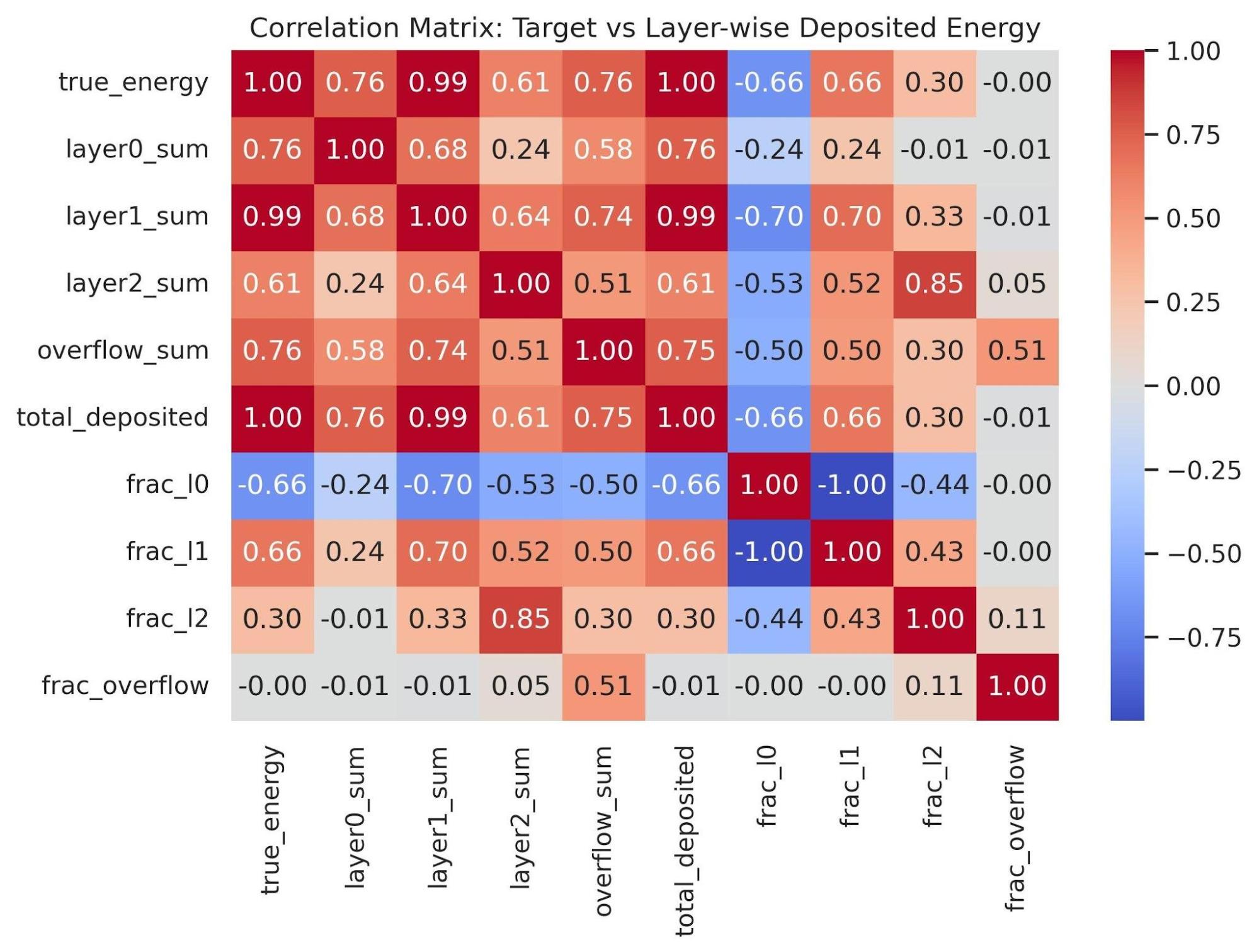}
\caption{Pearson correlation matrix of the engineered features against the true energy, highlighting the near-unit correlation of total deposited energy and the layer-fraction anticorrelations.}\label{fig:pearson}
\end{figure}

An exploratory analysis confirmed that several engineered observables track the incident energy very closely, as depicted in the Pearson correlation matrix in Fig.~\ref{fig:pearson}. The total deposited energy is almost perfectly correlated with the true energy ($r\approx1.00$), and the layer~1 energy is close behind ($r\approx0.99$), so a large part of the problem is already captured by simple global observables. The strong anticorrelation between the longitudinal energy fractions of different layers reflects energy conservation as deposition shifts with depth, which is the physical motivation for the sequence-based models. Any gain beyond a simple regression must therefore come from modelling shower fluctuations, leakage and higher-order correlations rather than the energy scale itself.

\subsection{Classical models}\label{sec:classres}

Table~\ref{tab:master} collects the reconstruction metrics. Although every model reaches a high coefficient of determination, the more sensitive metrics separate them clearly. The LSTM on longitudinal sequences gives the strongest classical result, with an RMSE of $0.429\pm0.052$~GeV and a resolution near 7.3\%, which shows that the ordered evolution of deposition is highly informative. The engineered-feature ANN follows closely at $0.442\pm0.040$~GeV, since the handcrafted variables already expose the relevant shower characteristics. The SVR reaches the lowest MAE (0.117~GeV) but a larger RMSE and worse resolution, a sign of sensitivity to outliers and leakage. The image-based CNN is the weakest classical model, at 7.972~GeV, which indicates that spatial information alone, without explicit longitudinal development, is not sufficient for accurate reconstruction.

\begin{table}[t]
\centering
\caption{Master regression performance. Multi-seed entries are mean~$\pm$~standard deviation; $\sigma_{\mathrm{rel}}$ is the energy resolution. The best overall model is the QLSTM.}
\label{tab:master}
\begin{tabular}{lcccccc}
\toprule
Model & $N$ & MAE & RMSE & $R^2$ & MRE & $\sigma_{\mathrm{rel}}$ \\
\midrule
Linear (trivial)   & 70k  & $\sim$0.10 & $\sim$0.14 & $\sim$0.999 & --    & --    \\
SVR (engineered)   & 70k  & 0.117 & 0.799 & 0.9992 & 0.005 & 0.203 \\
ANN (engineered)   & 70k  & $0.263\pm0.049$ & $0.442\pm0.040$ & 0.9998 & 0.011 & 0.093 \\
ANN (flattened)    & 70k  & 0.859 & 1.344 & 0.9978 & 0.072 & 0.276 \\
LSTM (full)        & 70k  & 0.210 & $0.429\pm0.052$ & 0.9997 & 0.008 & 0.073 \\
LSTM (matched)     & 1.5k & $0.488\pm0.037$ & $0.621\pm0.057$ & 0.9995 & 0.005 & 0.052 \\
CNN (multibranch)  & 70k  & 7.091 & 7.959 & 0.9227 & -0.138 & 0.098 \\
QNN (PCA-8)        & 2k   & 2.233 & 3.169 & 0.9888 & 0.032 & 0.175 \\
Quantum kernel     & 300  & 5.450 & 7.930 & 0.9145 & 0.320 & 1.222 \\
QCNN (hybrid)      & 1.5k & 1.545 & 2.441 & 0.9924 & 0.042 & 0.321 \\
QLSTM (hybrid)     & 1.5k & \textbf{0.160} & \textbf{0.294} & 0.9999 & 0.003 & 0.030 \\
\bottomrule
\end{tabular}
\end{table}

%\needinfo{In table 6, the resolution of LSTM matched (0.052) is better than LSTM full (0.073) in spite of its worst  RMSE; Can you please verify and reconcile these numbers?}

\begin{figure}
\centering
    \includegraphics[width=0.48\textwidth]{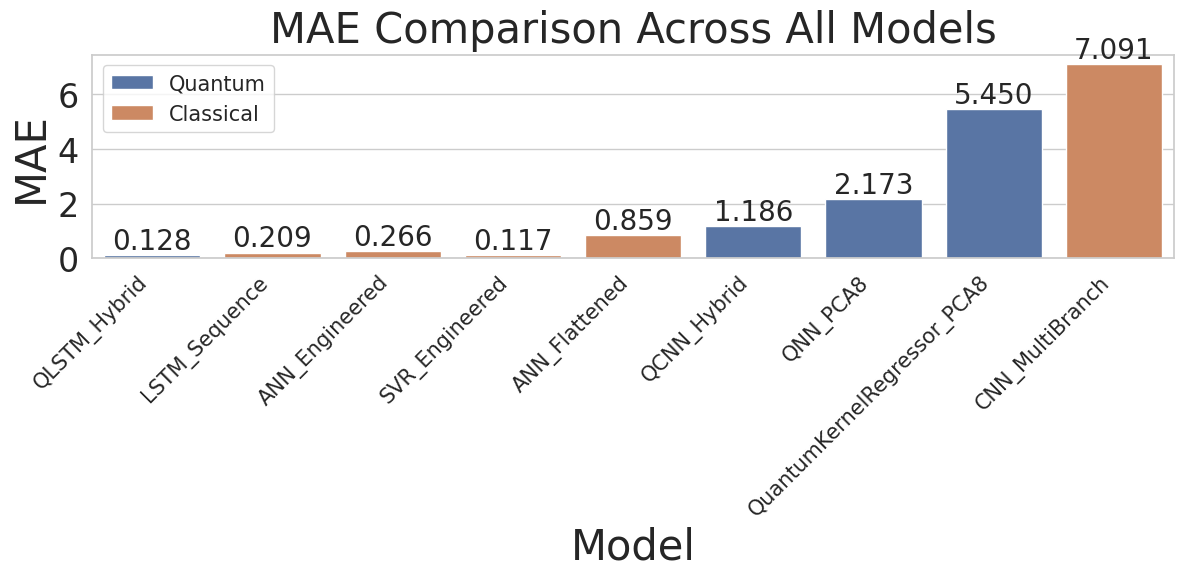}
    \includegraphics[width=0.48\textwidth]{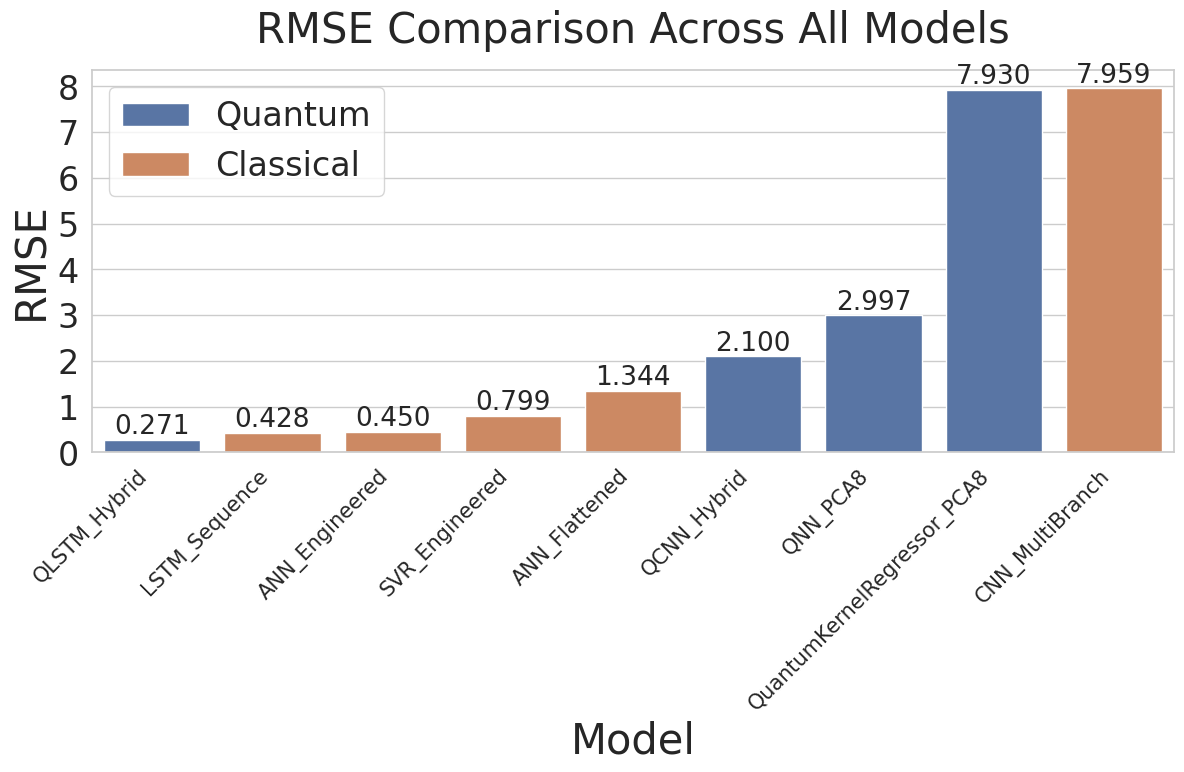}
    \includegraphics[width=0.60\textwidth]{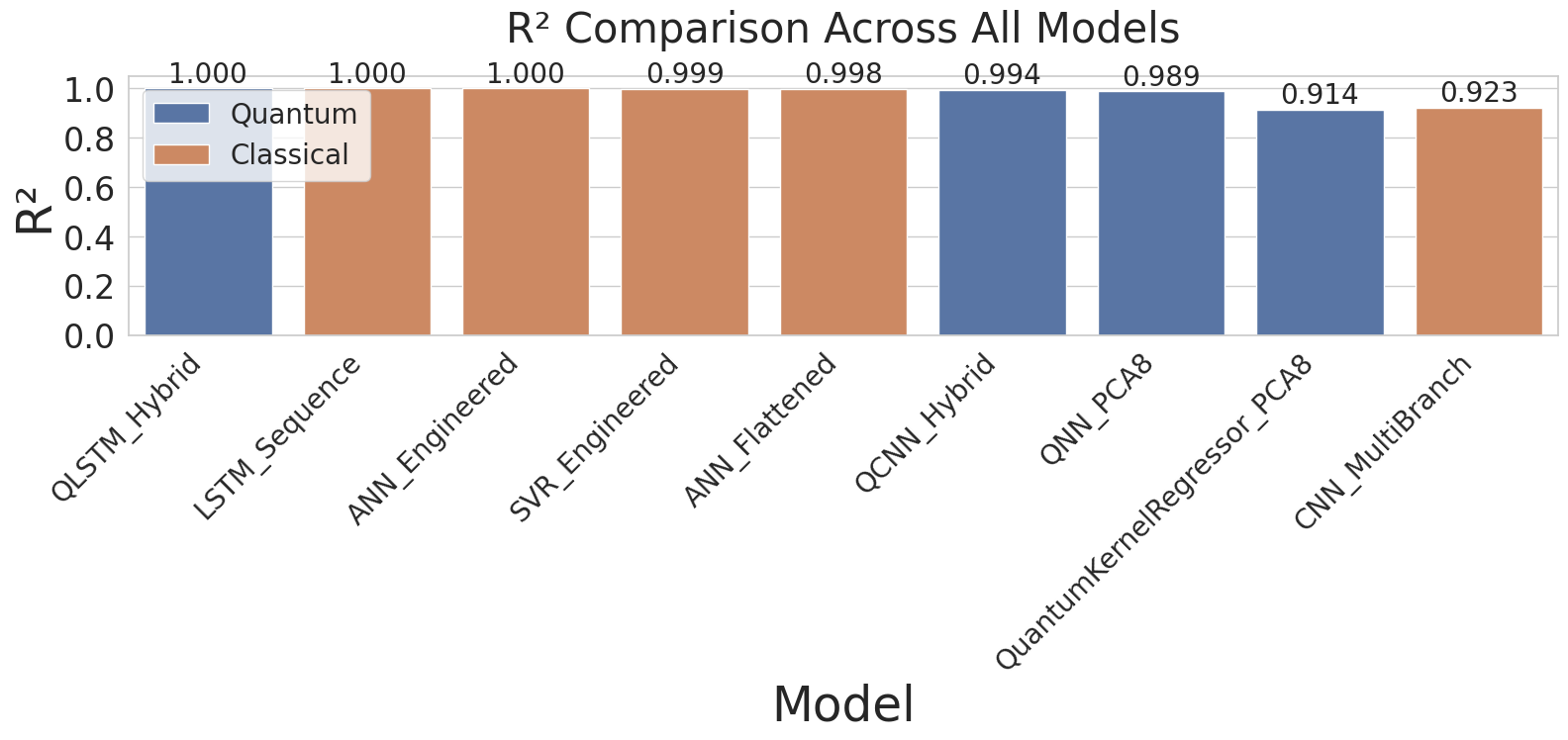}

\caption{Overall comparison of the evaluated models using (upper left) MAE, (upper right) RMSE, and (lower) $R^2$. While most models achieve coefficients of determination close to unity, the MAE and RMSE metrics provide a clearer separation of reconstruction performance, highlighting the superiority of the hybrid QLSTM.}
\label{fig:overallmetrics}
\end{figure}

\begin{figure}
\centering
\includegraphics[width=0.7\textwidth]{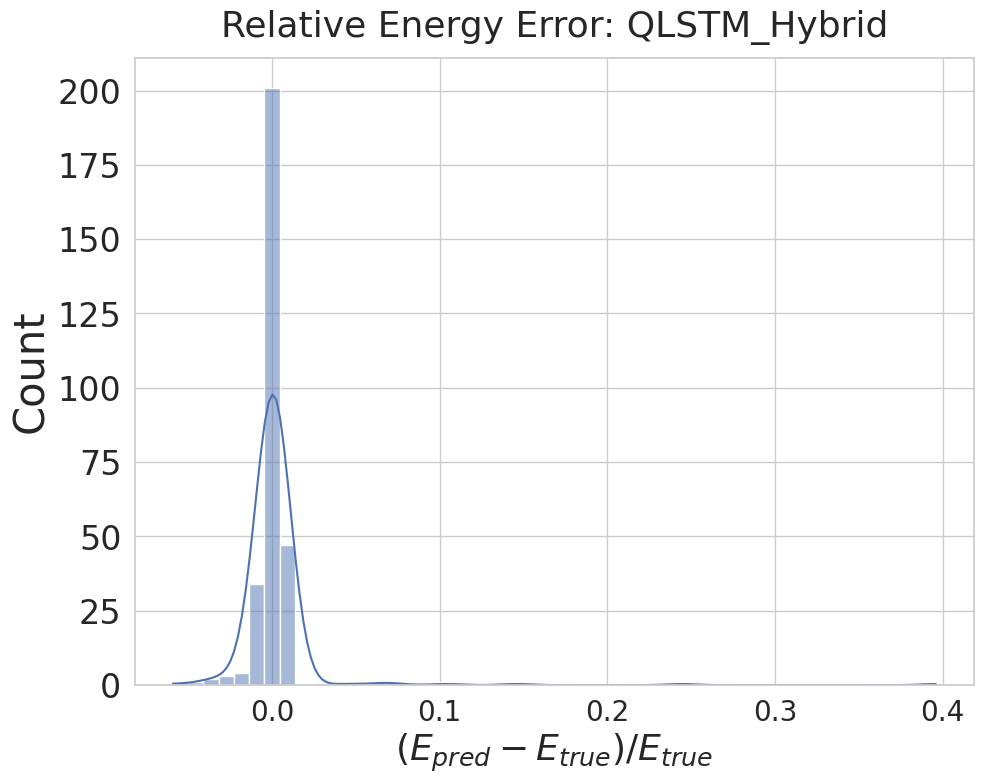}
\caption{Relative energy error $(\hat{E}-E)/E$ for the strongest classical models. The LSTM shows the sharpest, most symmetric peak; the engineered ANN is similar; the SVR distribution is wider with a positive tail.}\label{fig:releclass}
\end{figure}

A trend runs through the classical results: models built on physically informed representations beat those that read high-dimensional images, even with lower-dimensional inputs. The engineered-feature ANN and the sequence-based LSTM both outperform the image CNN, which suggests that the choice of representation weighs more than the raw complexity of the architecture. Figure~\ref{fig:overallmetrics} shows a classical comparison of MAE, RMSE and $R^2$, whereas Fig.~\ref{fig:releclass} shows the relative energy error for the strongest classical models.

%\begin{figure}
%\includegraphics[width=0.48\textwidth]{figures/MAE_CLASSICAL_MODELS.png}
%\includegraphics[width=0.48\textwidth]{figures/RMSE_CLASSICAL_MODELS.png}
%\caption{Classical comparison of MAE (left) and RMSE (right). SVR gives the lowest MAE; LSTM and the engineered ANN lead on RMSE; the multi-branch CNN is the weakest deep model.} \label{fig:classcomp}
%\end{figure}

\subsection{Quantum and hybrid models}\label{sec:quantres}

\begin{figure}
\centering
\includegraphics[width=0.48\textwidth]{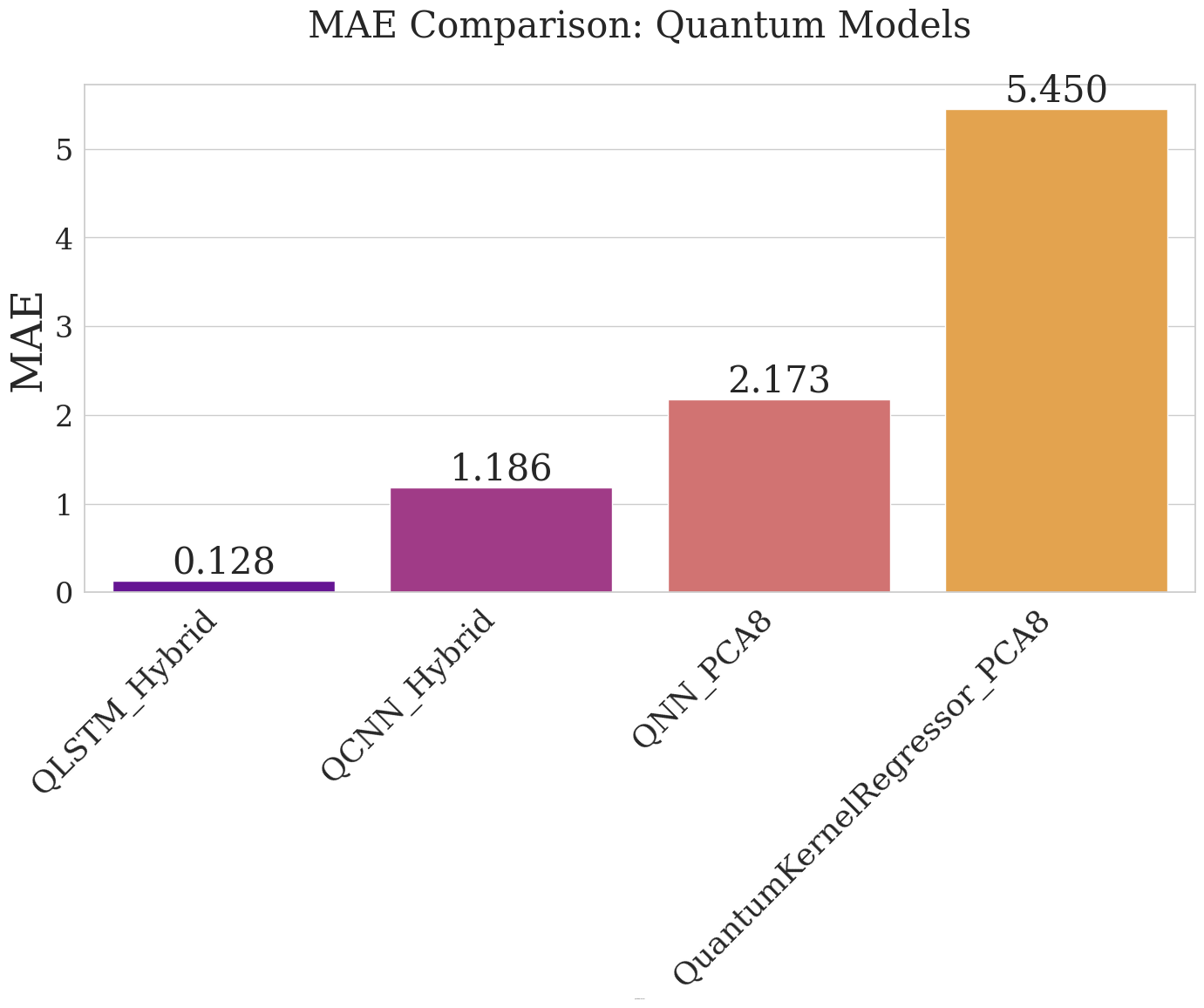}
\includegraphics[width=0.48\textwidth]{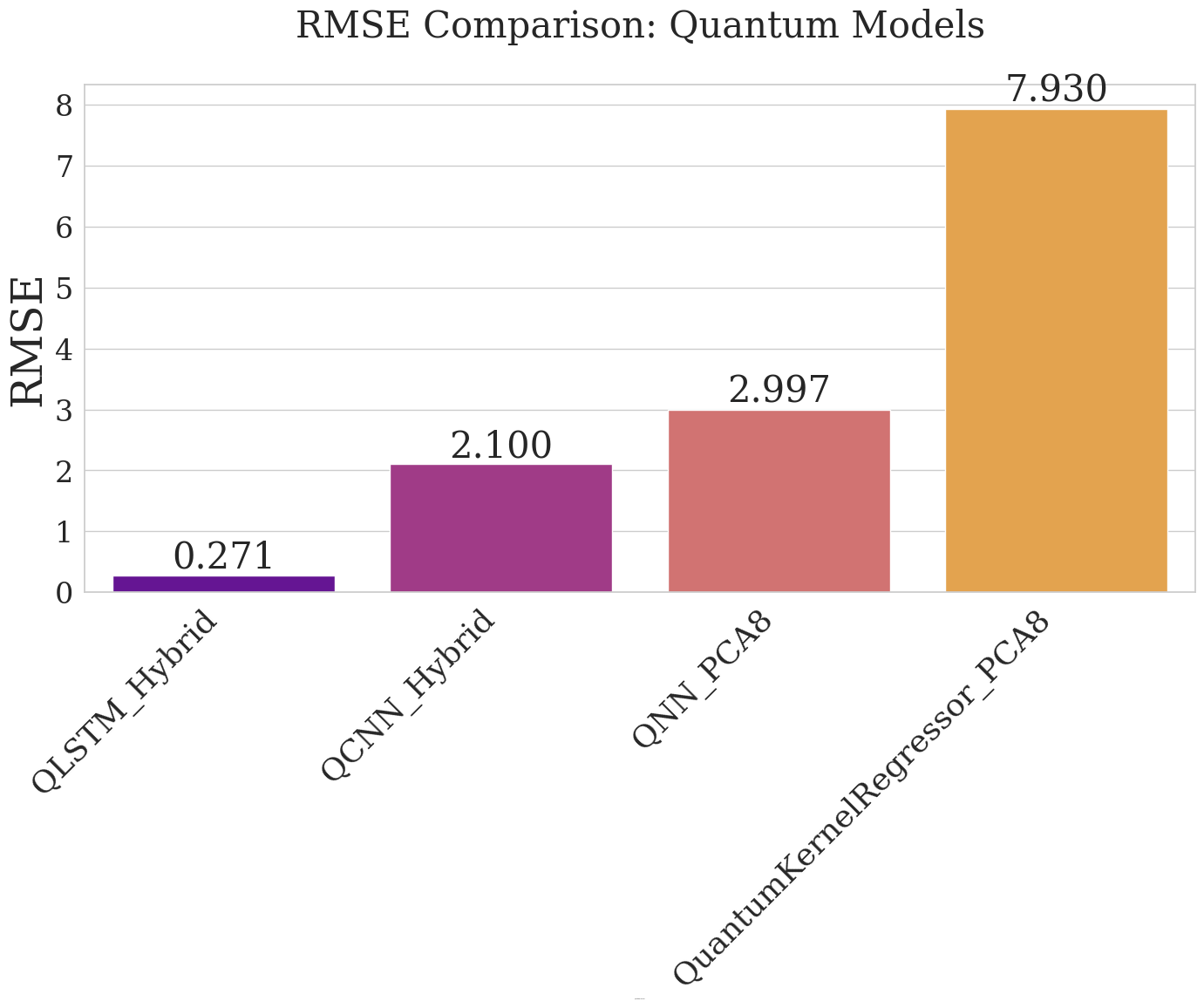}
\caption{Quantum comparison of MAE and RMSE. The QLSTM dominates; the quantum kernel is the weakest; the spread within the quantum family is wide.}
\label{fig:compq}
\end{figure}

\begin{figure}
\centering
\includegraphics[width=0.43\textwidth]{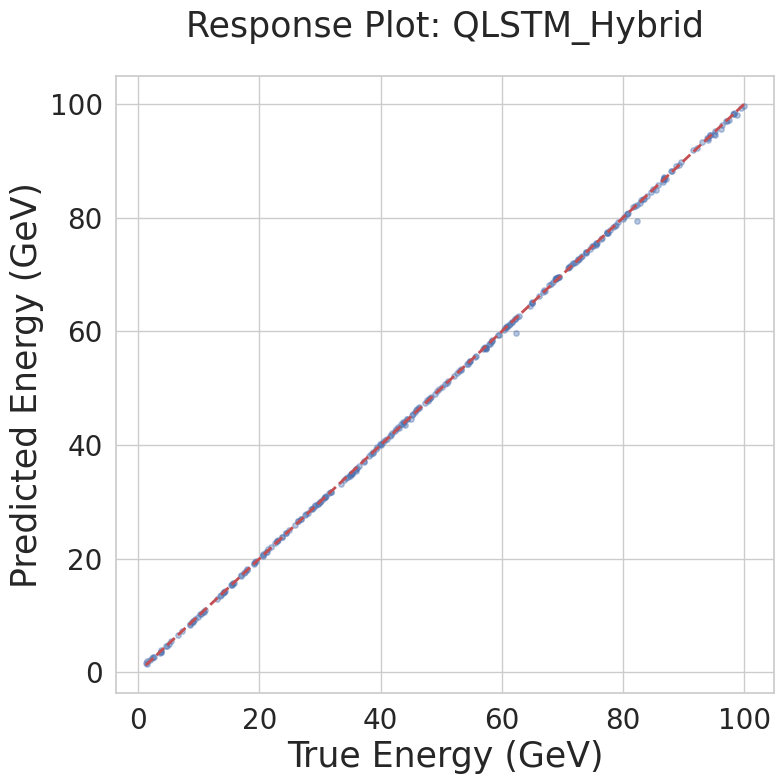}
\includegraphics[width=0.48\textwidth]{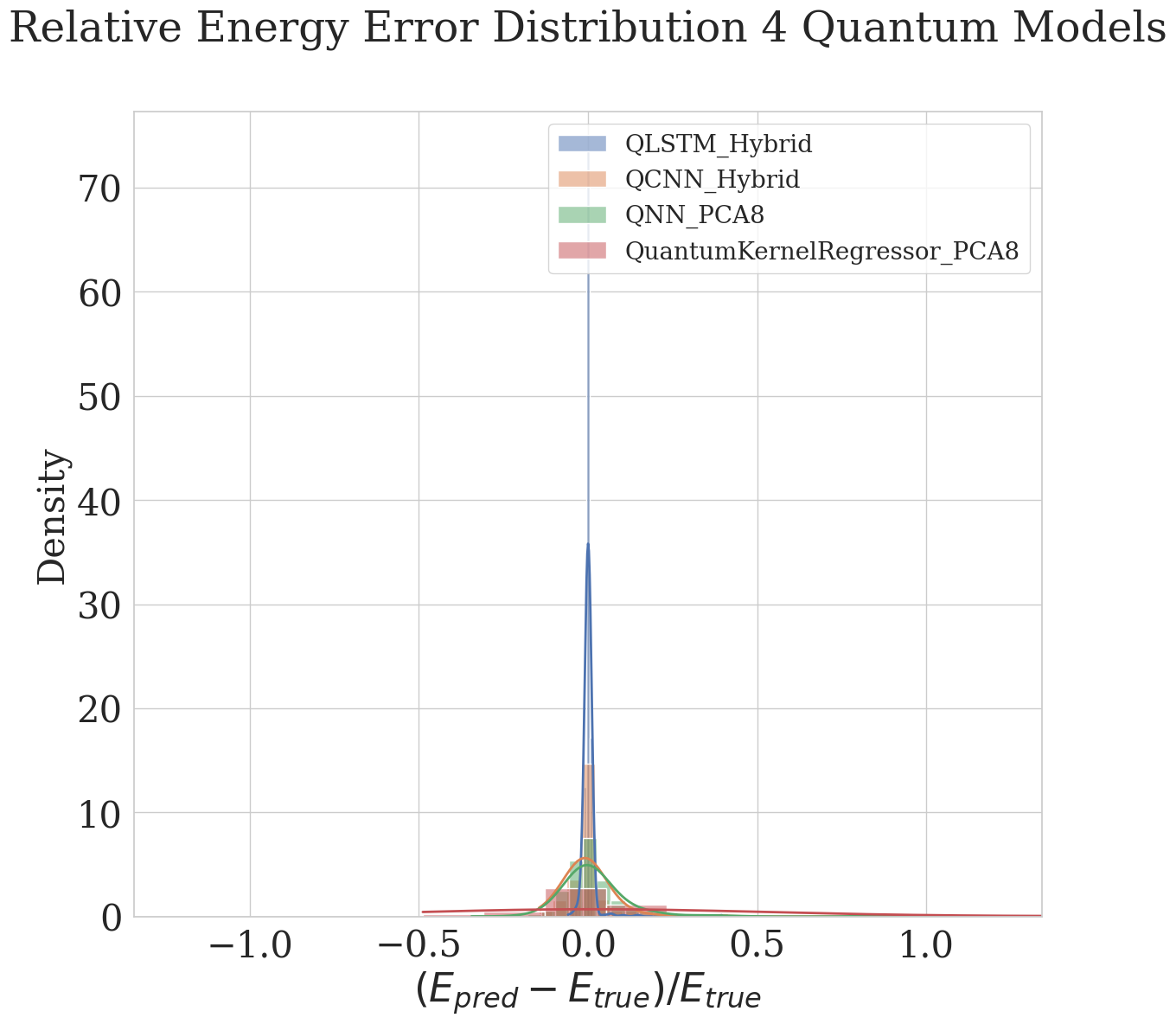}
\caption{(left) QLSTM response, reconstructed versus true energy across 1--100~GeV, tracking the identity line with $\sigma_{\mathrm{rel}}=3.0\%$ and only a few outliers near saturation; (right) combined relative-error distributions for all quantum models.}\label{fig:respq}
\end{figure}

The quantum models span more than an order of magnitude in performance, which again points to the architecture and the representation rather than to quantum computation alone. The QLSTM is the best of all eleven configurations, with an MAE of 0.160~GeV, an RMSE of 0.294~GeV and a resolution of 3.0\%, reached with only 1{,}500 training events. At the other end, the quantum kernel is the weakest, with an RMSE of 7.930~GeV and a resolution above 100\%, and the QNN also trails the best classical models, as shown in Fig.~\ref{fig:compq}. The QCNN sits in between, clearly above the classical CNN yet below the sequence-based models. The poor results of the quantum kernel and the QNN follow largely from the heavy dimensionality reduction they require: the PCA-8 input retains under a quarter of the variance, and the $O(N^2)$ kernel cost caps the quantum kernel at 300 training events. The QLSTM, by contrast, reads the longitudinal sequence directly and keeps the physical structure of the cascade. The QLSTM response and combined relative error distributions for all quantum models are depicted in Fig.~\ref{fig:respq}.

\subsection{Sample-efficiency analysis}\label{sec:effres}

Under the matched-subset protocol, all classical models lose accuracy when restricted to 1{,}500 events. The LSTM remains the strongest classical model but its RMSE rises from $0.429\pm0.052$~GeV on the full set to $0.621\pm0.057$~GeV. The QLSTM, trained on the same 1{,}500 events and the same representation, reaches 0.294~GeV, which lowers the error by about 52.7\% relative to the matched LSTM and also improves on the full-data LSTM. Because both factors, the representation and the training-set size, are held fixed in this comparison, the gain points to a more expressive latent representation in the hybrid model.

These results should be read with care. They rest on a single simulated dataset, one detector representation and idealized statevector simulation, and they do not establish a universal quantum advantage. With those limits stated, the matched-subset analysis offers preliminary evidence that hybrid quantum-classical sequence models can be more sample-efficient for calorimeter energy regression, which is a practical concern wherever labelled detector data are costly to generate.

\subsection{Physics-oriented energy resolution}\label{sec:resol}

\begin{figure}
\centering\includegraphics[width=0.7\textwidth]{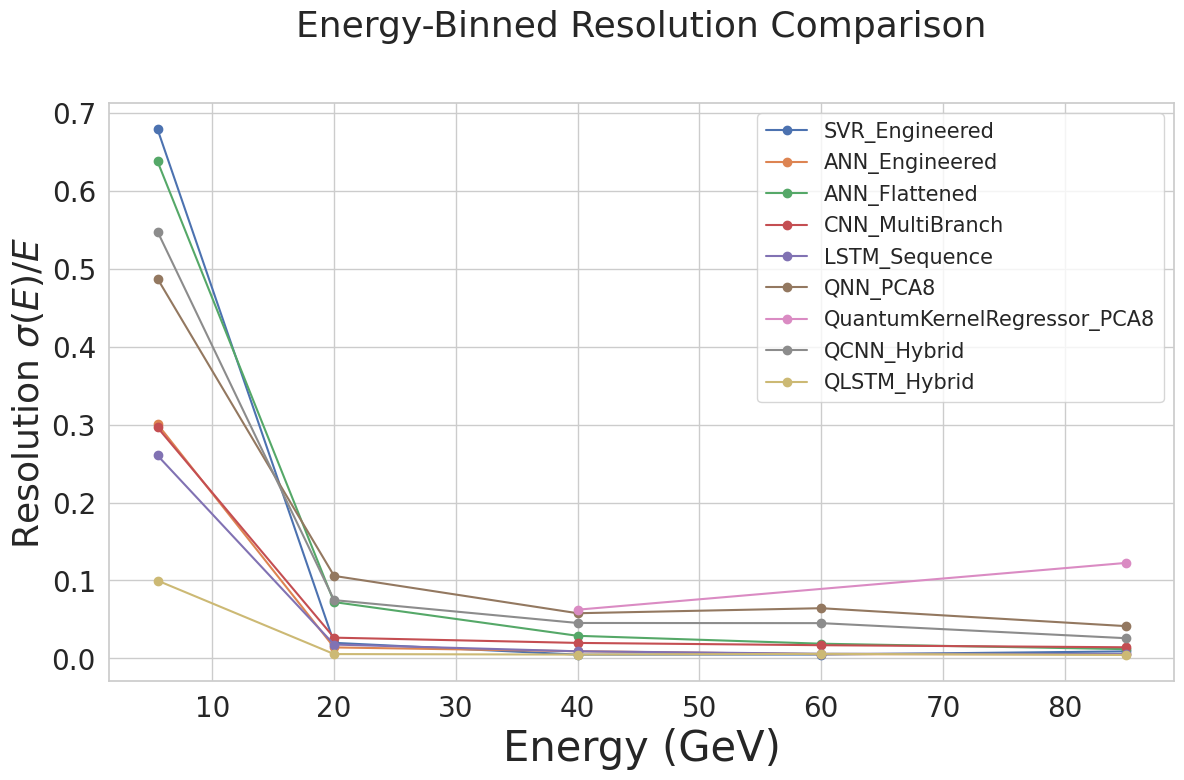}
\caption{Energy-binned resolution $\sigma(E)/E$ for all models. The QLSTM curve is the flattest and lowest; the sequence-based models follow the $1/\sqrt{E}$ stochastic scaling.}\label{fig:eneres}
\end{figure}

Evaluating the resolution $\sigma(E)/E$ as a function of incident energy (Fig.~\ref{fig:eneres}) connects the models to a quantity the calorimetry community uses directly. All models show larger relative uncertainty at low energy that improves as energy grows, consistent with the stochastic scaling of Eq.~\eqref{eq:stochastic}. The QLSTM gives the lowest and flattest curve, reaching about 3\% near 20~GeV and below 2\% at higher energies, with the classical LSTM second. Both sequence-based models reproduce the expected $1/\sqrt{E}$ behaviour, which is further evidence that the longitudinal representation preserves the information tied to deposition fluctuations. The CNN and the quantum kernel stay well above the rest across the range.

\subsection{Discussion}\label{sec:discussion}

The clearest message is that the representation of the detector response shapes reconstruction more than the choice between classical and quantum learning. Across nearly all models, physically informed encodings win, and the longitudinal sequence is the most informative description studied here: both the classical LSTM and the hybrid QLSTM lead their families while reading lower-dimensional inputs than the image models. The dominant difficulty in this problem is not extracting ever finer spatial patterns but modelling how deposition evolves with depth.

For the quantum side the picture is nuanced. The weak results of the quantum kernel and the QNN show that quantum computation does not by itself improve accuracy; aggressive compression and limited input information constrain them more than the quantum representation helps. The QLSTM, on the other hand, gives the best result overall, and its success appears to come from combining a meaningful sequential representation with a hybrid circuit able to learn within it. The matched-subset comparison is the strongest evidence for this reading, since the QLSTM beats the LSTM at equal data and equal representation. The advantage is reported with the parameter counts still missing and a single seed for the hybrid model, so the claim is promising rather than settled.

Several limitations remain. The analysis uses simulated data and ideal statevector simulation rather than real quantum hardware, and it covers one detector representation and one reconstruction task. Whether the trends survive realistic detector effects, noisy devices and other geometries is left for future work, which should also report the quantum parameter counts and multi-seed statistics needed to make the sample-efficiency claim conclusive.

%======================================================================
\section{Conclusions}\label{sec:conclusions}

This work compared classical and hybrid quantum-classical models for reconstructing incident positron energy from electromagnetic calorimeter data, across three detector representations and under controlled conditions. The representation of the calorimeter response emerged as the strongest factor: models that preserve the longitudinal development of the shower consistently outperform those based on engineered vectors or raw images. Among the classical models the LSTM led, which confirms that recurrent architectures suit the sequential nature of shower development.

Within the quantum and hybrid models, performance varied widely with the representation and the architecture. The quantum kernel and the QNN, which rely on heavily compressed inputs, showed limited capability, while the QCNN improved on the classical CNN yet stayed below the sequence models. The central result is the performance of the hybrid QLSTM, which reached the lowest reconstruction error of all models, an RMSE of 0.294~GeV, trained on only 1{,}500 events; under matched-data conditions it improved on the classical LSTM by about 52.7\% and also surpassed an LSTM trained on the full dataset. The resolution analysis supports this, with the sequence-based models following the expected stochastic scaling and the QLSTM giving the lowest relative resolution.

These findings do not amount to a universal quantum advantage, and they rest on simulated data, ideal statevector simulation and a single calorimeter configuration. Larger datasets, realistic detector effects, noisy hardware and alternative geometries are needed before practical deployment. What the study does show is that physically informed representations matter more than the choice between classical and quantum computation, and that hybrid quantum-classical sequence models are a promising direction for sample-efficient calorimeter reconstruction.

\section*{Data and code availability}
The CaloGAN dataset is publicly available~\cite{paganini2018calogan,nachman2017data}. %\needinfo{Add the link to the repository}

\section*{Acknowledgements}
%\needinfo{
AR acknowledges support from CIC-UMSNH (Mexico) under grant 18371. Also the authors extend their appreciation to the Deanship of Research and Graduate Studies at King Khalid University for funding this work through Large Research Project under grant number RGP2/155/46.%}

\bibliography{references}

\end{document}